\documentclass{jfm}
\usepackage{graphicx}
\usepackage{epstopdf}
\usepackage{float}
\usepackage{placeins}
\usepackage[table]{xcolor}
\usepackage{url}
\usepackage{tabularx}

\usepackage{amsmath}
\DeclareMathOperator{\sgn}{sgn}

\begin{document}

{ 
\shorttitle{Flow organization in laterally unconfined RB turbulence} 
\shortauthor{A. Blass {et al.}} 

\title{Flow organization in laterally unconfined Rayleigh-B\'enard turbulence}

\author
{
Alexander Blass\aff{1},
Roberto Verzicco\aff{2,1,3},\\
Detlef Lohse\aff{1,4},
Richard J.A.M. Stevens\aff{1}, 
\and 
Dominik Krug\aff{1}
\corresp{\email{d.j.krug@utwente.nl}},
}

\affiliation
{
\aff{1}
Physics of Fluids Group, Max Planck Center for Complex Fluid Dynamics, J. M. Burgers Center for Fluid Dynamics and MESA+ Research Institute, Department of Science and Technology, University of Twente, P.O. Box 217, 7500 AE Enschede, The Netherlands
\aff{2}
Dipartimento di Ingegneria Industriale, University of Rome "Tor Vergata". Via del Politecnico 1, Roma 00133, Italy
\aff{3}
Gran Sasso Science Institute -- Viale F. Crispi, 7 67100 L'Aquila, Italy.
\aff{4}
Max Planck Institute for Dynamics and Self--Organization, Am Fassberg 17, 37077 G\"ottingen, Germany
}

\maketitle
}

\begin{abstract}
We investigate the large-scale circulation (LSC) of turbulent Rayleigh-B\'enard convection in a large box of aspect ratio $\Gamma =32$ for Rayleigh numbers up to $Ra=10^9$ and at a fixed Prandtl number $Pr=1$. A  conditional averaging technique allows us to extract statistics of the LSC even though the number and the orientation of the structures vary throughout the domain.  We find that various properties of the LSC obtained here, such as the wall-shear stress distribution, the boundary layer thicknesses and the wind Reynolds number, do not differ significantly from results in confined domains ($\Gamma \approx 1$). This is remarkable given that the size of the structures (as measured by the width of a single convection roll) more than doubles at the highest $Ra$ as the confinement is removed. An extrapolation towards the critical shear Reynolds number of $Re_s^{\textrm{crit}} \approx 420$, at which the boundary layer (BL) typically becomes turbulent, predicts that the transition to the ultimate regime is expected at $Ra_{\textrm{crit}} \approx \mathcal{O}(10^{15})$ in unconfined geometries. This result is in line with the G\"ottingen experimental observations. Furthermore, we confirm that the local heat transport close to the wall is highest in the plume impacting region, where the thermal BL is thinnest, and lowest in the plume emitting region, where the thermal BL is thickest. This trend, however, weakens with increasing $Ra$.

\end{abstract} 

\section{Introduction}
Rayleigh-B\'enard (RB) convection \citep{ahl09,loh10,chi12,xia13} is the flow in a box heated from below and cooled from above. Such buoyancy driven flow is the paradigmatic example for  natural convection which often occurs in nature, e.g. in the atmosphere. For that case, a large-scale horizontal flow organization is observed in satellite pictures of weather patterns. Other examples include the thermohaline circulation in the oceans \citep{rah00}, the large-scale flow patterns that are formed in the outer core of the Earth \citep{gla99b}, where reversals of the large-scale convection roll are of prime importance, convection in gaseous giant planets \citep{bus94} and in the outer layer of the Sun \citep{mie00}. Thus, the problem is of interest in a wide range of scientific disciplines, including geophysics, oceanography, climatology, and astrophysics. 

For a given aspect ratio and given geometry, the dynamics in RB convection are determined by the Rayleigh number $Ra=\beta g\Delta H^3 /(\kappa \nu)$ and the Prandtl number $Pr=\nu/\kappa$. Here, $\beta$ is the thermal expansion coefficient, $g$ the gravitational acceleration, $\Delta$ the temperature difference between the horizontal plates, which are separated by a distance $H$, and $\nu$ and $\kappa$ are the kinematic viscosity and thermal diffusivity, respectively. The dimensionless heat transfer, i.e. the Nusselt number $Nu$, along with the Reynolds number $Re$ are the most important response parameters of the system. 

For sufficiently high $Ra$, the flow becomes turbulent, which means that there are vigorous temperature and velocity fluctuations. Nevertheless, a large-scale circulation (LSC) develops in the domain such that, in addition to the thermal boundary layer (BL), a thin kinetic BL is formed to accomodate the no-slip boundary condition near both the bottom and top plates.  Properties of the LSC and the nature of the BLs are highly relevant to the theoretical description of the problem. In particular, the unifying theory of thermal convection \citep{gro00,gro01,gro11,ste13} states that the transition from the classical to the ultimate regime takes place when the kinetic BLs become turbulent. This transition is shear based and driven by the large-scale wind, underlying the importance of the LSC to the overall flow behavior.

So far, the LSC and BL properties have mainly been studied in small aspect ratio cells, typically for $\Gamma=1/2$ and $\Gamma=1$. Various studies have shown that the BLs indeed follow the laminar Prandtl-Blasius (PB) type predictions in the classical regime \citep{ahl09,zho10,zho10b,ste12,shi15,shi17}. Previous studies by, for example \cite{wag12} and \cite{sch16}, have used results from direct numerical simulations (DNS) in aspect ratio $\Gamma=1$ cells to study the properties of the BLs in detail. \cite{wag12} showed that an extrapolation of their data gives that for $Pr=0.786$ the critical shear Reynolds number of $420$ is reached at $Ra\approx1.2\times10^{14}$. 

Despite the wealth of studies in low aspect-ratio domains, many natural instances of thermal convection take place in very large aspect ratio systems, as mentioned above. Previous research has demonstrated that several flow properties are significantly different in such unconfined geometries. \citet{har03} and \citet{har08} performed DNS at $Ra=\mathcal{O}(10^7)$ and $\Gamma =20$. They observed large-scale structures by investigating the advective heat transport and found the most energetic wavelength of the LSC at $4H-7H$. Recently, DNS by \cite{ste18} for $\Gamma =128$ and $Ra=\mathcal{O}(10^7-10^9)$ also reported `superstructures' with wavelengths of $6-7$ times the distance between the plates. Similar findings were made by \citet{pan18} over a wide range of Prandtl numbers $0.005\leq Pr \leq 70$ and $Ra$ up to $10^7$. It was shown that the signatures of the LSC can be observed close to the wall, which \cite{par04} described as clustering of thermal plumes originating in the BL and assembling the LSC. \citet{kru20} showed that the presence of the LSC leads to a pronounced peak in the coherence spectrum of temperature and wall-normal velocity. Based on DNS at $\Gamma=32$ and $Ra=\mathcal{O}(10^5-10^9)$, they determined that the wavelength of this peak shifts from $\hat{l}/H \approx 4$ to $\hat{l}/H \approx 7$ as $Ra$ is increased. 
 
\cite{ste18} have shown that in periodic domains, the heat transport is maximum for $\Gamma=1$ and reduces with increasing aspect ratio up to $\Gamma \approx 4$ when the large-scale value is obtained. They also found that fluctuation-based Reynolds numbers depend on the aspect ratio of the cell. However, other than the structure size, it is mostly unclear how the large-scale flow organization and BL properties are affected by different geometries. Not only is the size of the LSC more than 2 times larger without confinement (note that $\hat{l}$ measures the size of two counter-rotating rolls combined), but also other effects, such as corner vortices, are absent in periodic domains. Therefore one would expect differences in wind properties and BL dynamics. It is the goal of this paper to investigate these differences. Doing so comes with significant practical difficulty due to the random orientation of a multitude of structures that are present in a large box. To overcome this, we adopt the conditional averaging technique that was devised in \cite{ber20} to reliably extract LSC features even under these circumstances. Details on this procedure are provided in section \S \ref{section_results} after a short description of the dataset in \S \ref{section_method}. Finally, in \S\ref{shear_nusselt} and \S \ref{section_BL} we present results on how superstructures affect the flow properties in comparison to the flow formed in a cylindrical $\Gamma=1$ domain \citep{wag12}  and summarize our findings in \S\ref{section_conclusions}.

\section{Numerical method} \label{section_method}
The data used in this manuscript have previously been presented by \cite{ste18} and \cite{kru20}. A summary of the most relevant quantities for this study can be found in Table \ref{tab:stats}; note that there and elsewhere we use the free-fall velocity $V_{ff} = \sqrt{g\beta H\Delta}$ as a reference scale. In the following, we briefly report details on the numerical method for completeness. We carried out periodic RB simulations by numerically solving the three-dimensional incompressible Navier-Stokes equations within the Boussinesq approximation. They read:
\begin{equation} 
\frac{\partial \boldsymbol{u}}{\partial t} + \boldsymbol{u} \bcdot \bnabla \boldsymbol{u} =-\bnabla P 
+ \nu 
 \nabla^2\boldsymbol{u}+\beta g \theta \hat{z},
\label{eqn:Navier}
\end{equation}
\begin{equation} 
\bnabla \bcdot \boldsymbol{u} =0,
\label{eqn:div}
\end{equation}
\begin{equation} 
\frac{\partial \theta}{\partial t} + \boldsymbol{u} \bcdot \bnabla \theta = 
\kappa
\nabla ^2 \theta. 
\label{eqn:temp}
\end{equation}
%
Here, $\boldsymbol{u}$ is the velocity vector, $\theta$ the temperature, and the kinematic pressure is denoted by $P$. The coordinate system is oriented such that the unit vector $\hat{z}$ points up in the wall-normal direction, while the horizontal directions are denoted by $x$ and $y$. We solve (\ref{eqn:Navier}) - (\ref{eqn:temp}) using AFiD, the second-order finite difference code developed by Verzicco and coworkers \citep{ver96,poe15c}. We use periodic boundary conditions and a uniform mesh in the horizontal direction and a clipped Chebyshev-type clustering towards the plates in the wall-normal direction. For validations of the code against other experimental and simulation data in the context of RB we refer to  \citet{ver96,ver97,ver03,ste10,koo18}.

The aspect ratio of our domain is $\Gamma =L/H=32$, where $L$ is the length of the two horizontal directions of the periodic domain. The used numerical resolution ensures that all important flow scales are properly resolved \citep{shi10,ste10}. 

In this manuscript, we define the decomposition of instantaneous quantities into their mean and fluctuations such that $\psi (x,y,z,t)= \Psi (z) + \psi ' (x,y,z,t)$, where $\Psi = \langle \psi (x,y,z,t) \rangle _{x,y,t}$ is the temporal and horizontal average and $\psi '$ the fluctuations with respect to this mean.

\begin{figure}
 \centering
 \includegraphics[width=\textwidth]{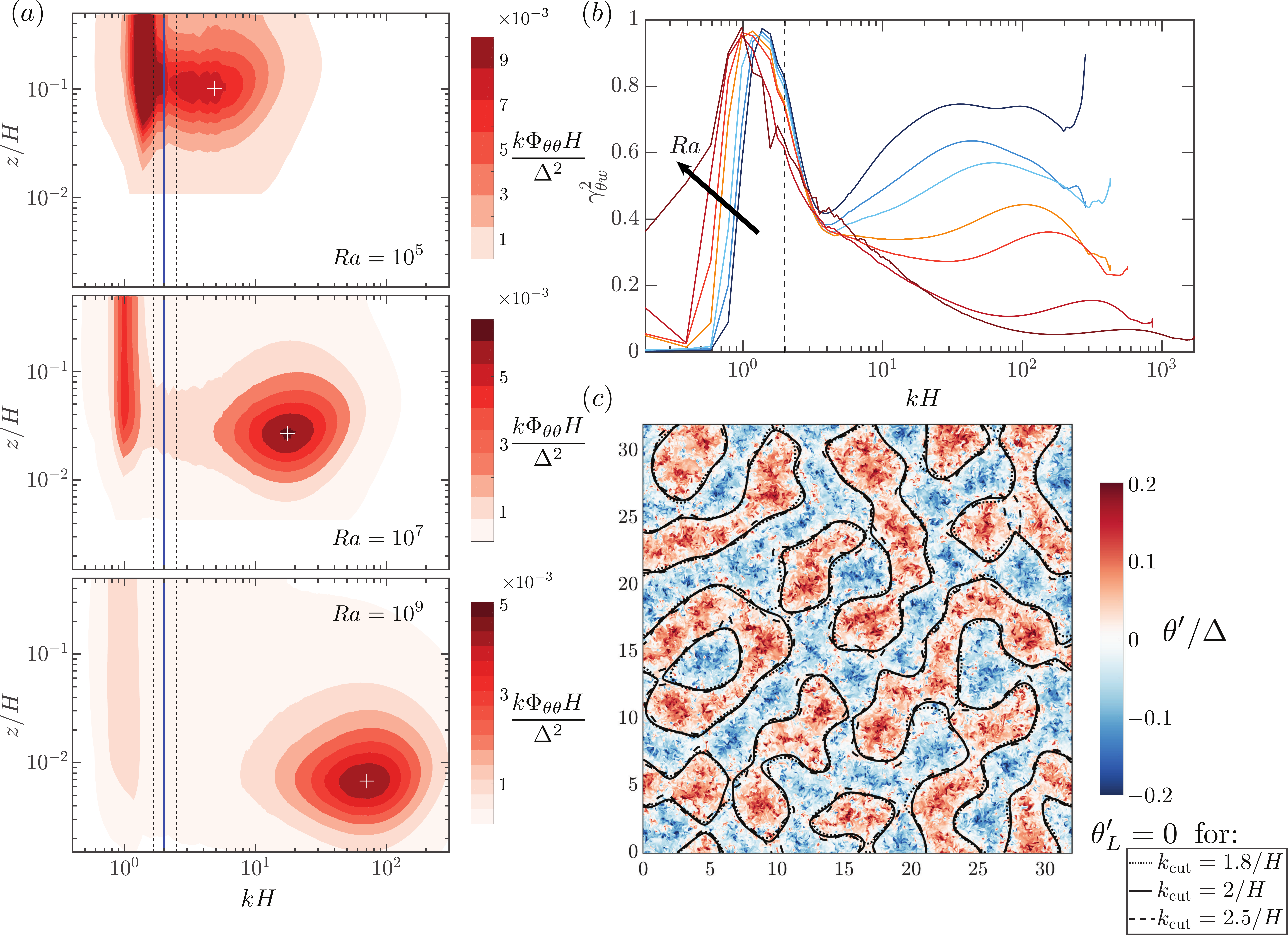}
 \caption{\label{fig:spectra} (a) Premultiplied temperature power spectra $k \Phi _{\theta \theta}$ for $Ra=10^5;10^7;10^9$. The blue line indicates the cut-off wavenumber $k_{\textrm{cut}} = 2/H$ used for the low-pass filtering. The dashed black lines indicate alternative cut-offs ($k_{\textrm{cut}} = 1.8/H $ and $k_{\textrm{cut}} = 2.5/H$) considered in panel (c). The white plusses are located at $k=0.57/\lambda _\theta ^*$ and $z=0.85 \lambda _\theta ^*$ (with $\lambda _\theta ^* =H/(2Nu)$) in all cases, which corresponds to the location of the inner peak \citep{kru20} (b) Coherence spectra of temperature and wall-normal velocity at mid-height, figure adopted from \cite{kru20}. The black line illustrates the choice of $k_{\textrm{cut}} = 2/H$ and the legend of figure \ref{fig:counts}a applies for the $Ra$-trend. (c) Snapshot of temperature fluctuations for $Ra=10^7$ at mid-height. The black lines show contours of $\theta_L ' =0$ evaluated for different choices of $k_{\textrm{cut}}$.}
\end{figure}

\begin{figure}
 \centering
 \includegraphics[width=\textwidth]{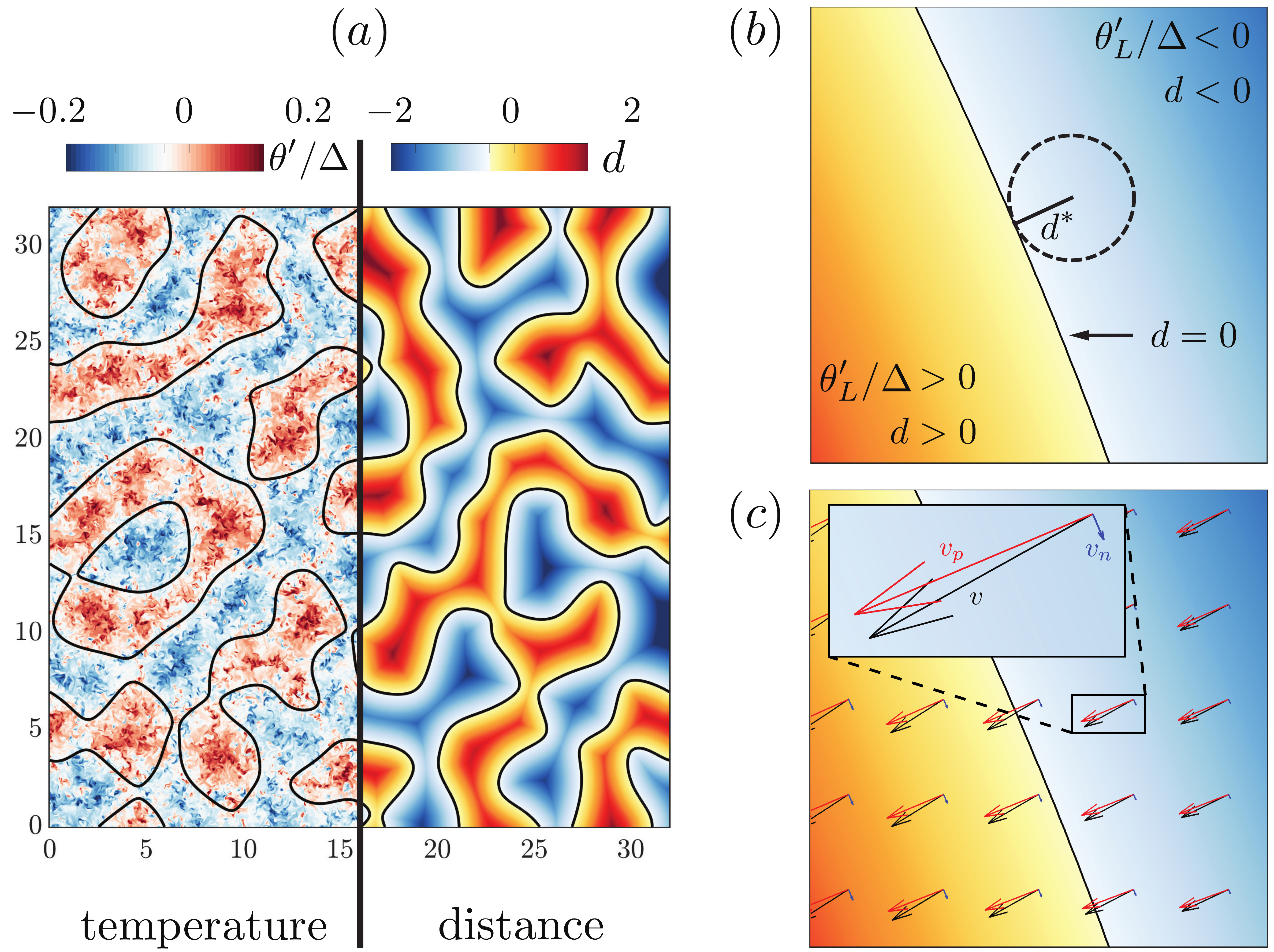}%
 \caption{\label{fig:method} Illustration of the conditional averaging method based on simulation data for $Ra=10^7$. (a) Temperature fluctuation field at mid-height and corresponding distance field (right). The black lines correspond to the zero-crossings $\theta _L' =0$ relative to which the distance $d^*$ is defined (see blow-up in panel b). Note that by definition isolines  $\theta _L '=0$   correspond to contours of $d=0$ in the distance field. (b) Illustration of the distance definition; for every point $d^*$ is equal to the radius of the smallest circle around that point which touches a $\theta _L' =0$ contour. (c) Illustration of the decomposition of the horizontal velocity $v$ into the parallel $v_p$ and the normal $v_n$ component to the gradient vector $d$. The color scheme in (b) and (c) indicates the $d$-field as in (a).}
\end{figure}
\begin{table} 
 \begin{center}
 \begin{tabular}{cccccc} 
 $Ra$ & $N_x \times N_y \times N_z$ & $Nu$ & $\hat{l}/H$ & $v_{RMS}/V_{ff}$ & $\lambda _\theta ^*/H$ \\[1pt]
 
 \hline \hline \\[-6.5pt]
 
 $ 1\times10^{5} $ & $2048 \times 2048 \times 64$ & 4.35 & 4.4 & 0.2172 & 0.115 \\
 $ 4\times10^{5} $ & $2048 \times 2048 \times 64$ & 6.48 & 4.5 & 0.2214 & 0.077 \\
 $ 1\times10^{6} $ & $3072 \times 3072 \times 96$ & 8.34 & 4.9 & 0.2198 & 0.060 \\
 $ 4\times10^{6} $ & $3072 \times 3072 \times 96$ & 12.27 & 5.4 & 0.2152 & 0.041 \\
 $ 1\times10^{7} $ & $4096 \times 4096 \times 128$ & 15.85 & 5.9 & 0.2107 & 0.032 \\
 $ 1\times10^{8} $ & $6144 \times 6144 \times 192$ & 30.94 & 6.3 & 0.1968 & 0.016 \\
 $ 1\times10^{9} $ & $12288 \times 12288 \times 384$ & 61.83 & 6.6 & 0.1805 & 0.008 \\ [1pt]
 
 \hline \hline 
 
 \end{tabular}
 \caption{Data from \cite{kru20} for the global Nusselt number, the grid resolution $(N_x,N_y,N_z)$ in streamwise, spanwise, and wall-normal direction, the location of the coherence spectra peak $\hat{l}$, the root mean square velocity $v_{RMS}=\sqrt{\langle v_x^2 + v_y^2 + w^2 \rangle _V}$ non-dimensionalized with the free-fall velocity $V_{ff}=\sqrt{\beta g H \Delta}$ and the estimated thermal BL thickness $\lambda_\theta ^*/H=1/(2Nu)$.}
 {\label{tab:stats}}
 \end{center}
\end{table}

\section{Conditional Averaging} \label{section_results}
Extracting features of the LSC in large aspect ratio cells poses a significant challenge. The reason is that there are multiple large-scale structures of varying sizes, orientation, and inter-connectivity at any given time. It is therefore not possible to extract properties of the LSC by using methods that rely on tracking a single or a fixed small number of convection cells, which have been proven to be successful in analyzing the flow in small \citep{sun08,wag12} to intermediate \citep{ree08} aspect-ratio domains. To overcome this issue, we use a conditional averaging technique developed in \cite{ber20}, where this framework was employed to study the modulation of small-scale turbulence by the large flow scales. This approach is based on the observation of \cite{kru20} that the premultiplied temperature power spectra $k\Phi_{\theta\theta}$ (shown in figure \ref{fig:spectra}a) is dominated by two very distinct contributions. One is due to the `superstructures' whose size (relative to $H$) increases with increasing $Ra$ and typically corresponds to wavenumbers $kH \approx 1-1.5$. The other contribution relates to a `near-wall peak' with significantly smaller structures whose size scales with the thickness of the BL \citep{kru20}. This implies that this peak shifts to larger $k$ (scaled with $H$) as the BLs get thinner at higher $Ra$. Hence, there is a clear spectral gap between superstructures and small-scale turbulence, which widens with increasing $Ra$, as can readily be seen from figure \ref{fig:spectra}a. This figure also demonstrates that a spectral cut-off $k_{\textrm{cut}} = 2/H $ is a good choice to separate superstructure contributions from the other scales over the full $Ra$ range $10^5\leq Ra\leq 10^9$ considered here.

The choice for $k_{\textrm{cut}} = 2/H $ is further supported by considering the spectral coherence 
\begin{equation}
 \gamma ^2 _{\theta w}(k)=\frac{|\Phi_{\theta w}(k)|^2}{\Phi_{\theta \theta}(k)\Phi_{w w}(k)},
\end{equation}
where $\Phi_{w w}$ and $\Phi_{\theta w}$ are the velocity power spectrum and the co-spectrum of $\theta$ and $w$, respectively. The coherence $\gamma^2$ may be interpreted as a measure of the correlation \emph{per scale}. The results at $z = 0.5H$ in figure \ref{fig:spectra}b indicate that there is an almost perfect correlation between $\theta'$ and $w'$ at the superstructure scale. Almost no energy resides at the scales corresponding to the high-wavenumber peak in $\gamma ^2 _{\theta w}$ \citep[see][]{kru20}, such that the coherence there is of little practical consequence. The threshold $k_{\textrm{cut}} = 2/H$ effectively delimits the large-scale peak in $\gamma^2_{\theta w}$ towards larger $k$ for all $Ra$ considered, such that this value indeed appears to be a solid choice to distinguish the large-scale convection rolls from the remaining turbulence. To confirm this, we overlay a snapshot of $\theta'$ with zero-crossings of the low-pass filtered signal (with cut-off wavenumber $k_{\textrm{cut}}$) $\theta'_L$ in figure \ref{fig:spectra}c. These contours reliably trace the visible structures in the temperature field. Furthermore, it becomes clear that slightly different choices for $k_{\textrm{cut}}$ do not influence the contours significantly. This is consistent with the fact that only limited energy resides at the scales around $k\approx 2/H$, such that $\theta'_L$ only changes minimally when $k_{\textrm{cut}}$ is varied within that range.  In the following, we adopt $k_{\textrm{cut}} = 2/H$ to obtain $\theta'_L$ except when we study the effect of the choice for $k_{\textrm{cut}}$. 

We use $\theta'_L$ evaluated at mid-height to map the horizontal field onto a new horizontal coordinate $d$. To obtain this coordinate, first the distance $d^*$ to the nearest zero-crossing in $\theta'_L$ is determined for each point in the plane. This can be achieved efficiently using a nearest-neighbor search. Then the sign of $d$ is determined by the sign of $\theta'_L$, such that $d$ is given by
\begin{equation}
d=\sgn(\theta '_L){d^*}.
\end{equation}
All results presented here are with reference to the lower hot plate. Hence $d<0$ and $d>0$ correspond to plume impacting and plume emitting regions, respectively. The averaging procedure is illustrated in figure \ref{fig:method}a,b. Another important aspect is a suitable decomposition of the horizontal velocity component $v$. Figure \ref{fig:method}c shows how we decompose $v$ into one component ($v_p$) parallel the local gradient $\nabla d$, and another component ($v_n$) normal to it. This ensures that $v_p$ is oriented normal to the zero-crossings in $\theta'_L$ for small $|d|$, where the wind is strongest. However, at larger $|d|$, the orientation may vary from a simple interface normal, which accounts for curvature in the contours. It should be noted that the $d$-field is determined at mid-height and consequently applied to determine the conditional average at all $z$-positions. This is justified since \citet{kru20} showed that there is a strong \emph{spatial} coherence of the large scales in the vertical direction. Therefore, the resulting zero-contours would almost be congruent if $\theta'_L$ was evaluated at other heights. The time-averaged conditional average is obtained by averaging over points of constant $d$, while we make use of the symmetry around the mid-plane to increase the statistical convergence. Mathematically, the conditioned averaging results in a triple decomposition according to $\psi (x,y,z,t) = \Psi (z) + \overline{\psi}(z,d)+ \widetilde{\psi}(x,y,z,t)$, where the overline indicates conditional and temporal averaging.

\begin{figure}
 \centering
 \makebox[0pt]{\includegraphics[width=1.13\textwidth]{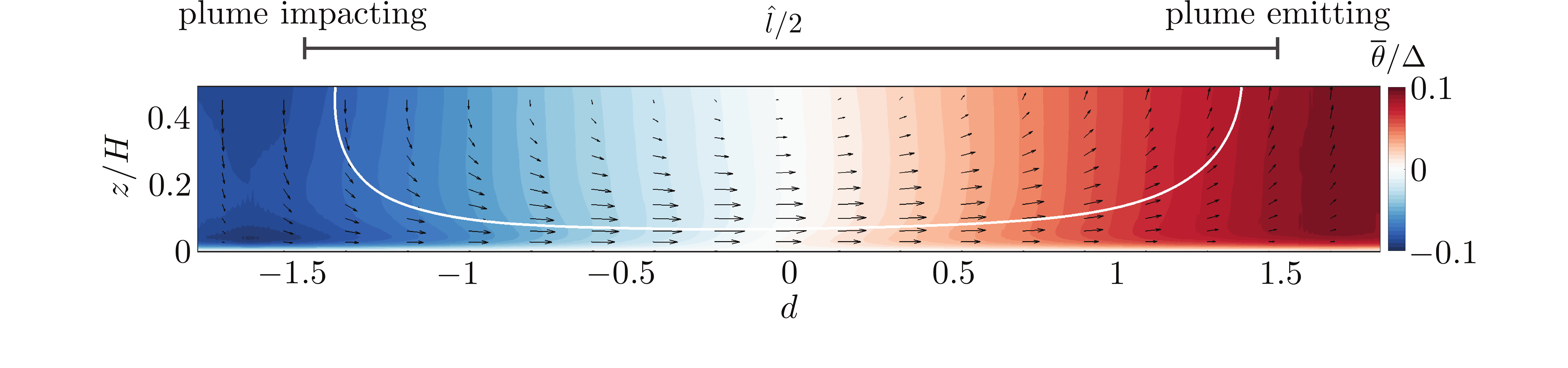}}
 \caption{\label{fig:lsc} Contour plot of the conditionally averaged temperature $\overline{\theta}/\Delta$ for $Ra=10^7$. The arrows show $\overline{w}/V_{ff}$ and $\overline{v}_p/V_{ff}$ and are plotted every 24 and every 6 data points along $d$ and $z$, respectively. The white line is the streamline which passes through $z^*/H$ at $d=0$.}
\end{figure}

Applying the outlined method to our RB dataset results in a representative large-scale structure like the one depicted in figure \ref{fig:lsc} for $Ra= 10^7$. In general, we find $\overline{\theta}<0$ with predominantly downward flow for $d<0$, while lateral flow towards increasing $d$ dominates in the vicinity of $d = 0$. In the plume emitting region $d>0$ the conditioned temperature $\overline{\theta}$ is positive and the flow upward. In interpreting the results it is important to keep in mind that the averaging is `sharpest' close to the conditioning location ($d = 0$) and `smears out' towards larger $|d|$ as the size of individual structures varies.  We normalize $d$ with $\hat{l}$ to enable a comparison of results across $Ra$. Based on the location of the peak in $\gamma^2$, \cite{kru20} found that the superstructure size is $\hat{l}=5.9 H$ at $Ra= 10^7$. As indicated, the conditionally averaged flow field in figure \ref{fig:lsc} corresponds to approximately half this size. 

\begin{figure}
 \centering
 \includegraphics[width=\textwidth]{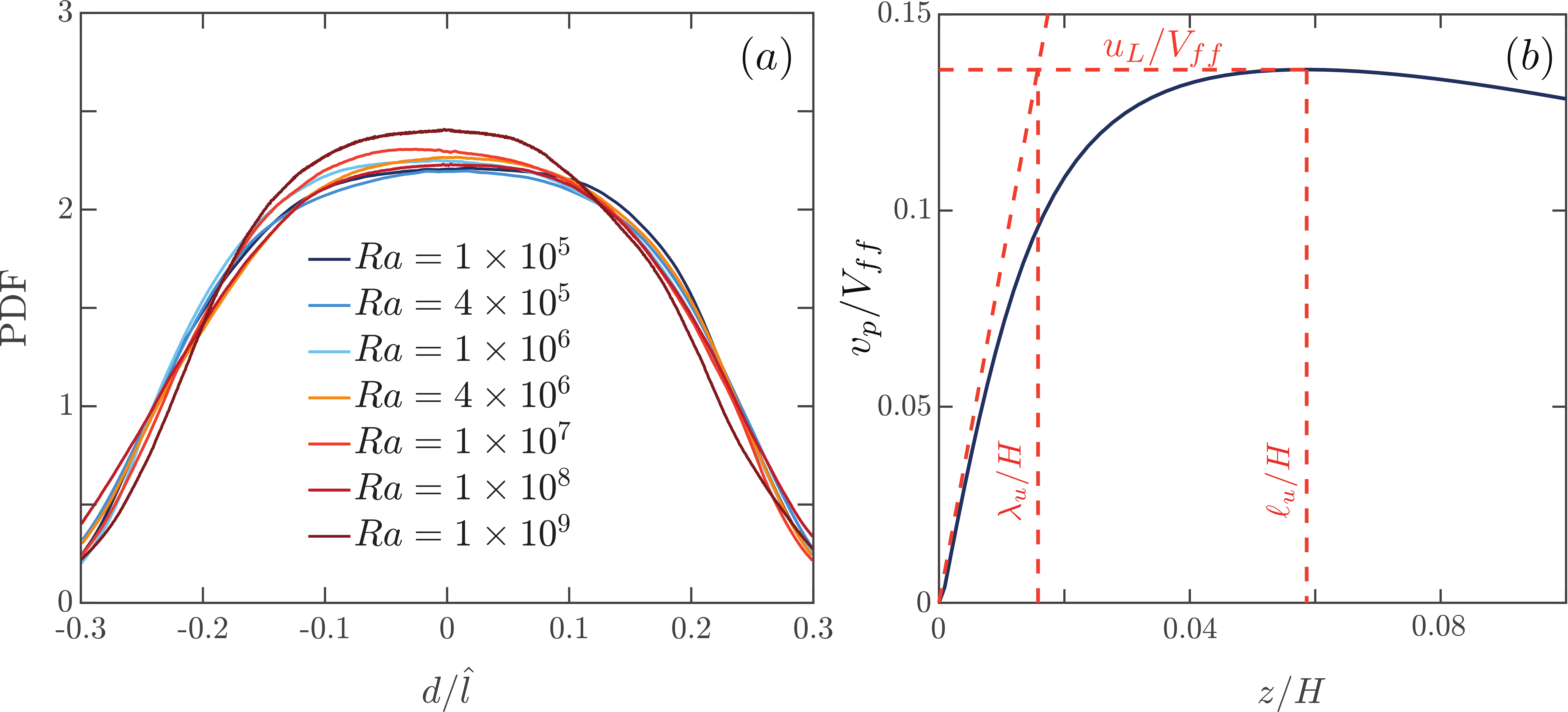}%
 \caption{\label{fig:counts} (a) PDF of the normalized distance parameter $d/\hat{l}$. (b) Sample velocity profile to illustrate  the slope method ($\lambda$) and the level method ($\ell$) used to determine the instantaneous BL thicknesses.}
\end{figure}

We present the probability density function (PDF) of the distance parameter $d$ in figure \ref{fig:counts}a. The data collapse to a reasonable degree, indicating that there are no significant differences in how the LSC structures vary in time and space across the considered range of $Ra$. Visible deviations are at least in part related also to uncertainties in determining $\hat{l}$ via fitting the peak of the $\gamma^2$-curve. 

The LSC is carried by $v_p$, which is also supported by the fact that the velocity component normal to the gradient $\nabla d$ averages to zero, i.e. $\overline{v}_n \approx 0 $, for all $d$. The determination of the viscous BL thickness is therefore based on $v_p$ only. We use the `slope method' to determine the viscous ($\lambda _u$) and thermal ($\lambda_\theta$) BL thickness. Both are determined locally in space and time and are based on instantaneous wall-normal profiles of $\theta$ and $v_p$, respectively. As sketched in figure 4b, $\lambda$ is given by the location at which linear extrapolation using the wall-gradient reaches the level of the respective quantity. Here the `level' (e.g. $u_L$ for velocity) is defined as the local maximum within a search interval above the plate. In agreement with \cite{wag12} we find that the results for both thermal and viscous BL do not significantly depend on the search region when it is larger than $4 \lambda_\theta ^*$. Therefore, we have adopted this search region in all our analyses.

\begin{figure}
 \centering
 \includegraphics[width=\textwidth]{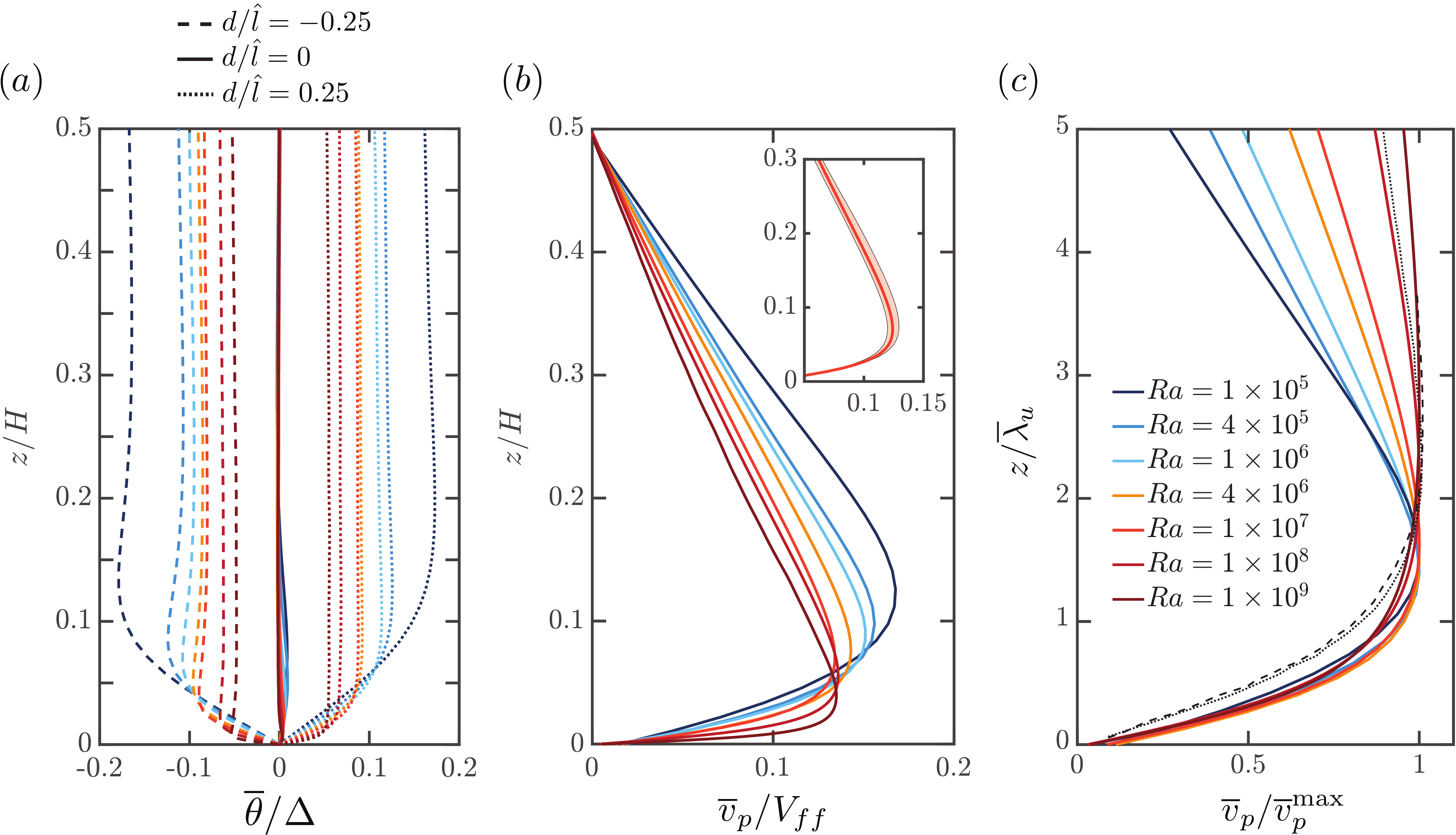}%
 \caption{\label{fig:T_vp} (a) Conditioned temperature $\overline{\theta}/\Delta$ at $d=0$ and in the plume impacting $(d/\hat{l}=-0.25)$ and in the plume emitting region $(d/\hat{l}=0.25)$ for various $Ra$, see legend in (c). (b) Wind velocity $\overline{v}_p/V_{ff}$ at $d=0$ versus $z/H$ at the same $Ra$. The inset shows the sensitivity of the results to different choices of $k_{\textrm{cut}}$ in the range $1.8\leq k_{\textrm{cut}} H \leq 2.5$ (same range used in figure \ref{fig:spectra}) for $Ra=10^7$. (c) Mean wind velocity normalized by its maximum value for various $Ra$ (see legend). The dashed and dotted black lines in (c) represent experimental data from \cite{sun08} at $\Gamma =1$ for $Ra=1.25 \times 10^9$ and $Ra=1.07 \times 10^{10}$, respectively.}
\end{figure}

In figure \ref{fig:T_vp}a we present the conditionally averaged temperature $ \overline{\theta}$ as a function of $z/H$ at three different locations of $d/\hat{l}$. Consistent with the conditioning on zero-crossings in $\theta'_L =0$, we find that $\overline{\theta}\approx 0$ for all $z$ at $d = 0$. In the plume impacting ($d/\hat{l} = -0.25$) and emitting ($d/\hat{l} = 0.25$) regions, $\overline{\theta}$ is respectively negative and positive throughout. On both sides, $\overline{\theta}$ attains nearly constant values in the bulk, the magnitude of which is decreasing significantly with increasing $Ra$.

Profiles for the mean wind velocity $\overline{v}_p(z) $ at $d = 0$ are shown in figure \ref{fig:T_vp}b,c. These figures show that the viscous BL becomes thinner with increasing $Ra$, while the decay from the velocity maximum to $0$ at $z/H = 0.5$ is almost linear for all cases. We note that of all presented results the wind profile is most sensitive to the choice of the threshold $k_{\textrm{cut}}$. The reason is that the obtained wind profile depends on both the contour location and orientation. To provide a sense for the variations associated with the choice of $k_{\textrm{cut}}$, we compare the present result at $Ra = 10^7$ to what is obtained using alternative choices ($k_{\textrm{cut}} = 1.8/H$ and $k_{\textrm{cut}} = 2.5/H$) in the inset of figure \ref{fig:T_vp}b. This plot shows that results within the BL are virtually insensitive to the choice of $k_{\textrm{cut}}$ while the differences in the bulk consistently remain below 5\%. In panel (c) of figure \ref{fig:T_vp} we re-plot the data from figure \ref{fig:T_vp}b normalized with the BL thickness $\overline{\lambda}_u(d=0)$ and the velocity maximum $\overline{v}_p^{\textrm{max}}$. The figure shows that the velocity profiles for the different $Ra$ collapse reasonably well for $z\lessapprox \overline{\lambda}_u$. A comparison to the experimental data by \cite{sun08}, which were recorded in the center of a slender box with $\Gamma =1$ and $Pr = 4.3$, reveals that, although the overall shape of the profiles is similar, there are considerable differences in the near-wall region. With their precise origin unknown, these discrepancies could be related to the differences in $Pr$ and $\Gamma$ but also to experimental uncertainties.

\begin{figure}
 \centering
 \includegraphics[width=\textwidth]{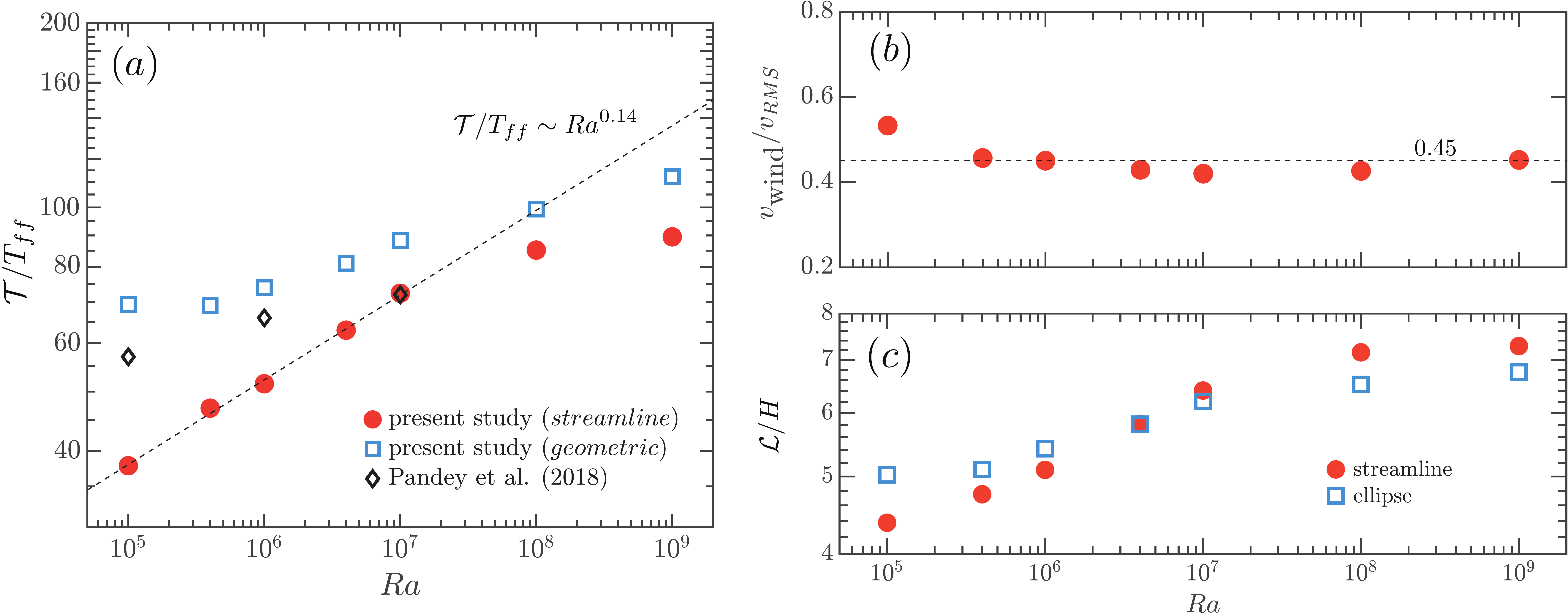}%
 \caption{\label{fig:timescale} (a) Timescale $\mathcal{T}$ versus $Ra$ using different methods. The datasets are: The time needed to circulate the flow along a streamline, which passes through $ z^*/H$ at $d=0$ (red circles), see figure \ref{fig:lsc}; the timescale calculated with the geometric method of \cite{pan18} (blue squares). We also show the \cite{pan18} data itself, which were calculate for the smaller  $Pr=0.7$ (black diamonds). (b) Average velocity $v_{\textrm{wind}}$ determined along the streamline chosen in (a), normalized with $v_{RMS}$. (c) Comparison between the length of the streamline and the circumference $\pi (0.25 \hat{l}+0.5 H)$ of the ellipse (geometric method), both used to calculate the respective timescale in (a).}
\end{figure}

Another interesting question that we can address based on our results concerns the evolution timescale $\mathcal{T}$ of the LSC. We estimate $\mathcal{T}$ as the time it takes a fluid parcel to complete a full cycle in the convection roll. To do this we compute the streamline that passes through the location $z^*/H$ of the velocity maximum $\overline{v}_p(z^*/H)=\overline{v}_p^{\max}$ at $d=0$ as shown in figure \ref{fig:lsc}. The  integrated travel time $\mathcal{T}$ along this averaged streamline as a function of $Ra$ is presented in figure \ref{fig:timescale}a. 
We find $\mathcal{T}/T_{ff} \gg 1$, i.e., 
the typical timescale of the LSC dynamics is much longer than the free-fall time $T_{ff} = \sqrt{H / (\beta g \Delta)}$. 
Up to $Ra =10^7$ the timescale $\mathcal{T}$ grows approximately according to $\mathcal{T}/T_{ff} 
\sim Ra^{0.14}$, but the trend flattens out at $Ra$ beyond that value.

To compare our results to other estimates in the literature, we also adopt the method used by \citet{pan18} to estimate $\mathcal{T}$. These authors assumed the LSC to be an ellipse and set the effective velocity to $\frac{1}{3} \, v_{RMS}$, where the prefactor $1/3$ is purely empirical. The results for the `geometric method' are compared to the corresponding results by \citet{pan18} in figure \ref{fig:timescale}a. Results are consistent between the two methods in terms of the order of magnitude. However, the actual values, especially at lower $Ra$, differ significantly, and also the trends do not fully agree. The streamline approach allows us to determine the average convection velocity along the streamline $v_{\textrm{wind}}\equiv \mathcal{L}/\mathcal{T}$, where $\mathcal{L}$ is the length of the streamline. Figure \ref{fig:timescale}b show that $v_\textrm{wind}$ is indeed proportional to $v_{RMS}$ with $v_\textrm{wind} \approx 0.45 \, v_{RMS}$ in the considered $Ra$ number regime. In figure \ref{fig:timescale}c, we present $\mathcal{L}$ along with the ellipsoidal estimate used in \citet{pan18}. From this, it appears that an ellipse does not very well represent the streamline geometry. Further, it becomes clear that it is the difference in the length-scale estimate that leads to the different scaling behaviors for $\mathcal{T}$ in figure \ref{fig:timescale}a. 

\begin{figure}
 \centering
 \includegraphics[width=\textwidth]{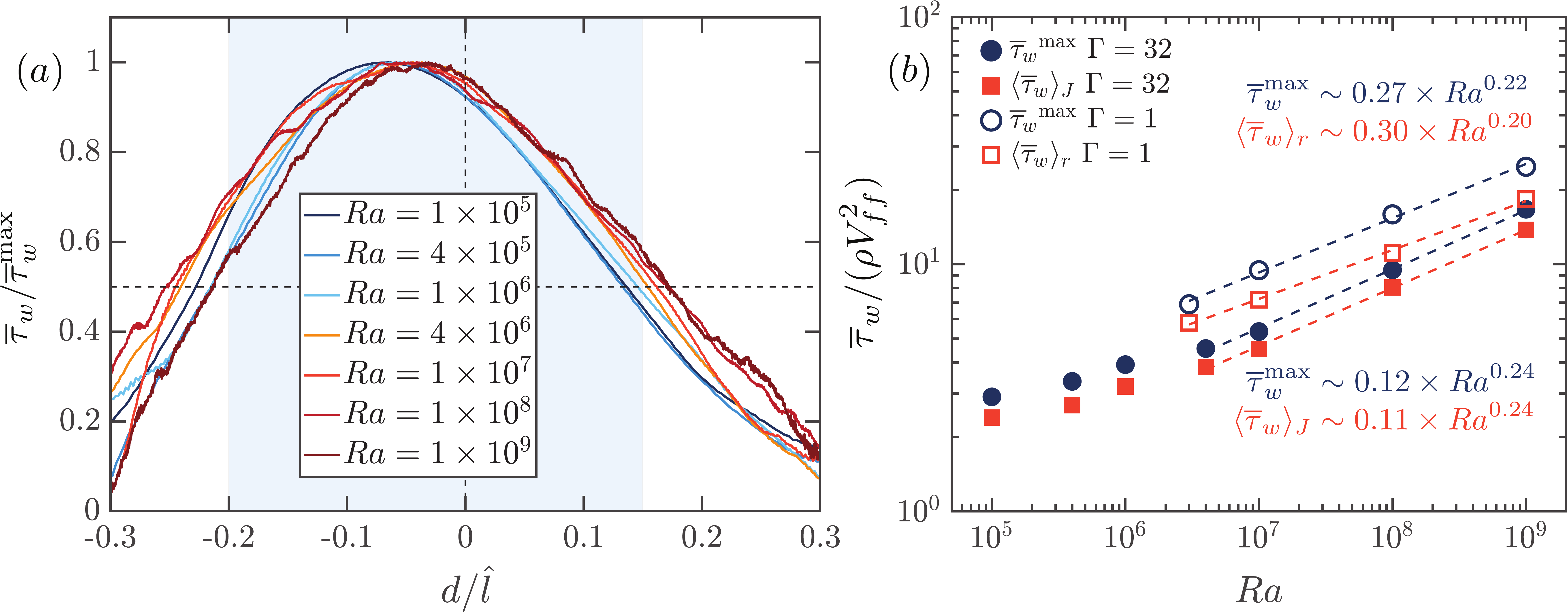}%
 \caption{\label{fig:tau_w}(a) Normalized shear stress $\overline{\tau} _w$ as a function of $d/\hat{l}$ and (b) mean shear stress $\langle \overline{\tau}_w \rangle _J$ and maximum shear stress $\overline{\tau}_w ^{\max}$ versus $Ra$. The filled symbols show data of the present study ($\Gamma=32$ periodic domain), while the open symbols represent the data of \cite{wag12} for $\Gamma=1$ with a cylindrical domain. The blue symbols show the maximum shear stress and the red symbols the mean shear stress over the interval $J=\{d/\hat{l} \; | \; d/\hat{l} \in[-0.2 : 0.15] \}$.}
\end{figure}

\section{Wall shear stress and heat transport}
\label{shear_nusselt}
The shear stress $\overline{\tau}_w $ at the plate surface is defined through
\begin{equation}
\overline{\tau} _w/\rho  = -  \nu 
\langle \partial _z \overline{v}_p \rangle _t.
\end{equation}
Here $\partial _z$ is the spatial derivative in wall-normal direction. In figure \ref{fig:tau_w}a we show that the normalized shear stress $ \overline{\tau}_w / \overline{\tau}_w ^{\max}$ as a function of the normalized distance $d/\hat{l}$ is nearly independent of $Ra$. Similar to findings in smaller cells \citep{wag12}, the curves are asymmetric with the maximum ($d/\hat{l} \approx -0.05$) shifted towards the plume impacting region. The value of $ \overline{\tau}_w / \overline{\tau}_w ^{\max}$ drops to about 0.25 in both the plume impacting ($d/\hat{l} = -0.25$) and the plume emitting region ($d/\hat{l} = 0.25$).

We use the good collapse of the $ \overline{\tau}_w / \overline{\tau}_w ^{\max}$ profiles across the full range of $Ra$ considere to separate regions with significant shear from those with little to no lateral mean flow. We define the `wind' region based on the approximate criterion $ \overline{\tau}_w / \overline{\tau}_w ^{\max} \gtrapprox 0.5$, which leads to the interval $J=\{d/\hat{l} \; | \; d/\hat{l} \in[-0.2 : 0.15] \}$ that is indicated by the blue shading in figure \ref{fig:tau_w}a. We use the average over this interval to evaluate the wind properties and indicate this by $\langle \rangle_J$. In figure \ref{fig:tau_w}b the data for mean $\langle \overline{\tau}_w\rangle_J$ and for maximum $\overline{\tau}_w ^{\max}$ wall shear stress are compiled for the full range of $Ra$ considered. Both quantities are seen to increase significantly as $Ra$ increases. Around $Ra=1\times 10^6\textup{--}4\times 10^6$ we can see a transition point at which the slope steepens. For lower $Ra$ the scaling of $\langle \overline{\tau} _w \rangle_{J}$ is much flatter. A fit to the data for $Ra\geq 4\times 10^6$ gives
\begin{equation}
{\overline{\tau}_w /\rho \over  V_{ff}^2} \sim Ra^{0.24},
\end{equation}
for both $\langle \overline{\tau}_w\rangle_J$ and $\overline{\tau}_w ^{\max}$. Overall, we find that the shear stress at the wall due to the turbulent thermal superstructures (in the periodic $\Gamma=32$ domain with $Pr=1$) compares well with the shear stress in a cylindrical $\Gamma =1$ domain by \cite{wag12} with $Pr=0.786$. Most importantly, the scaling with $Ra$ is the same for both cases. The actual shear stress seems to be somewhat higher in the cylindrical aspect ratio $\Gamma=1$ domain than in the periodic domain in which the flow is unconfined. In part this difference may be related to the difference in $Pr$. Besides that, as we will show in the next section, the shear Reynolds number is slightly lower for the periodic domain than in the confined domain.

\begin{figure}
 \centering
 \includegraphics[width=\textwidth]{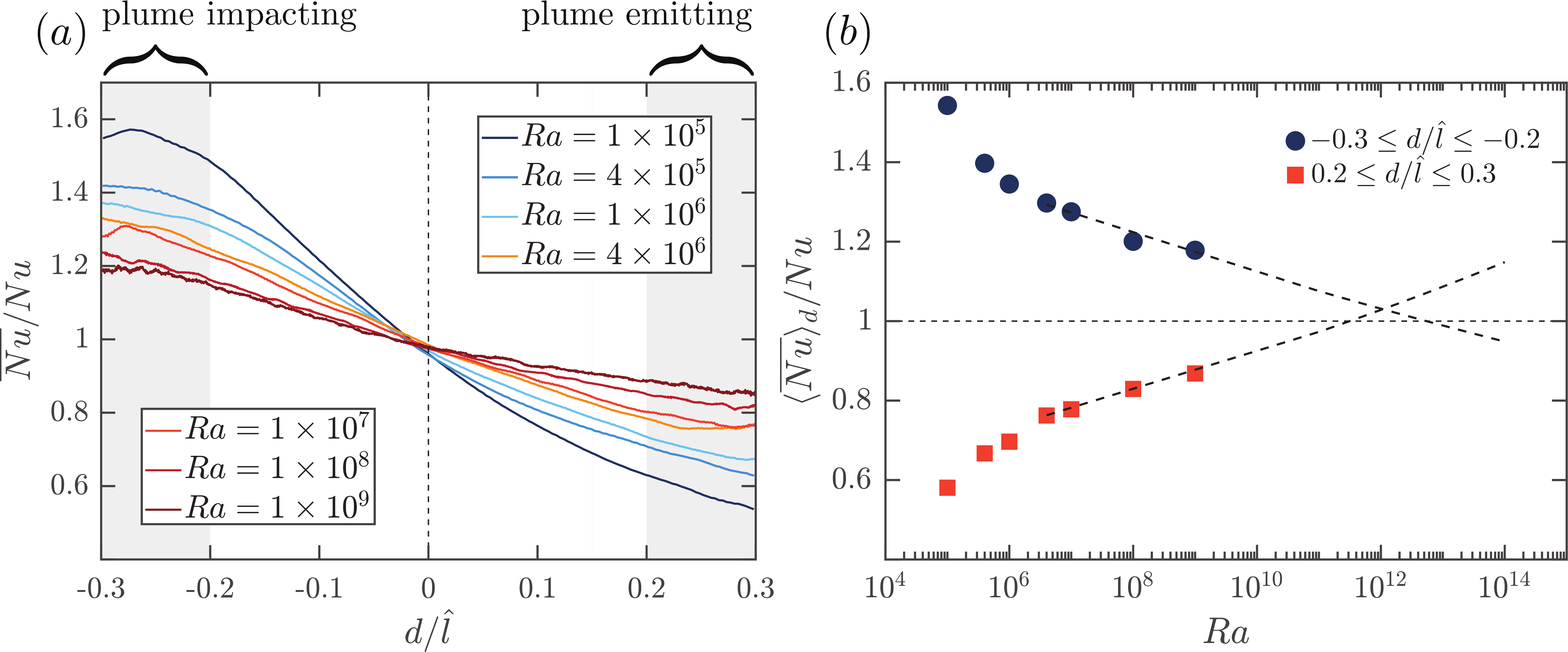}%
 \caption{\label{fig:Nu}(a) Local heat flux $\overline{Nu}$ at the wall normalized by the global heat flux $Nu$ as function of the normalized spatial variable $d/ \hat{l}$. (b) Values in the impacting ($ -0.3 \leq d/ \hat{l} \leq -0.2$) and emitting ($0.2 \leq d/ \hat{l} \leq 0.3$) range as a function of $Ra$.}
\end{figure}

\begin{figure}
 \centering
 \includegraphics[width=\textwidth]{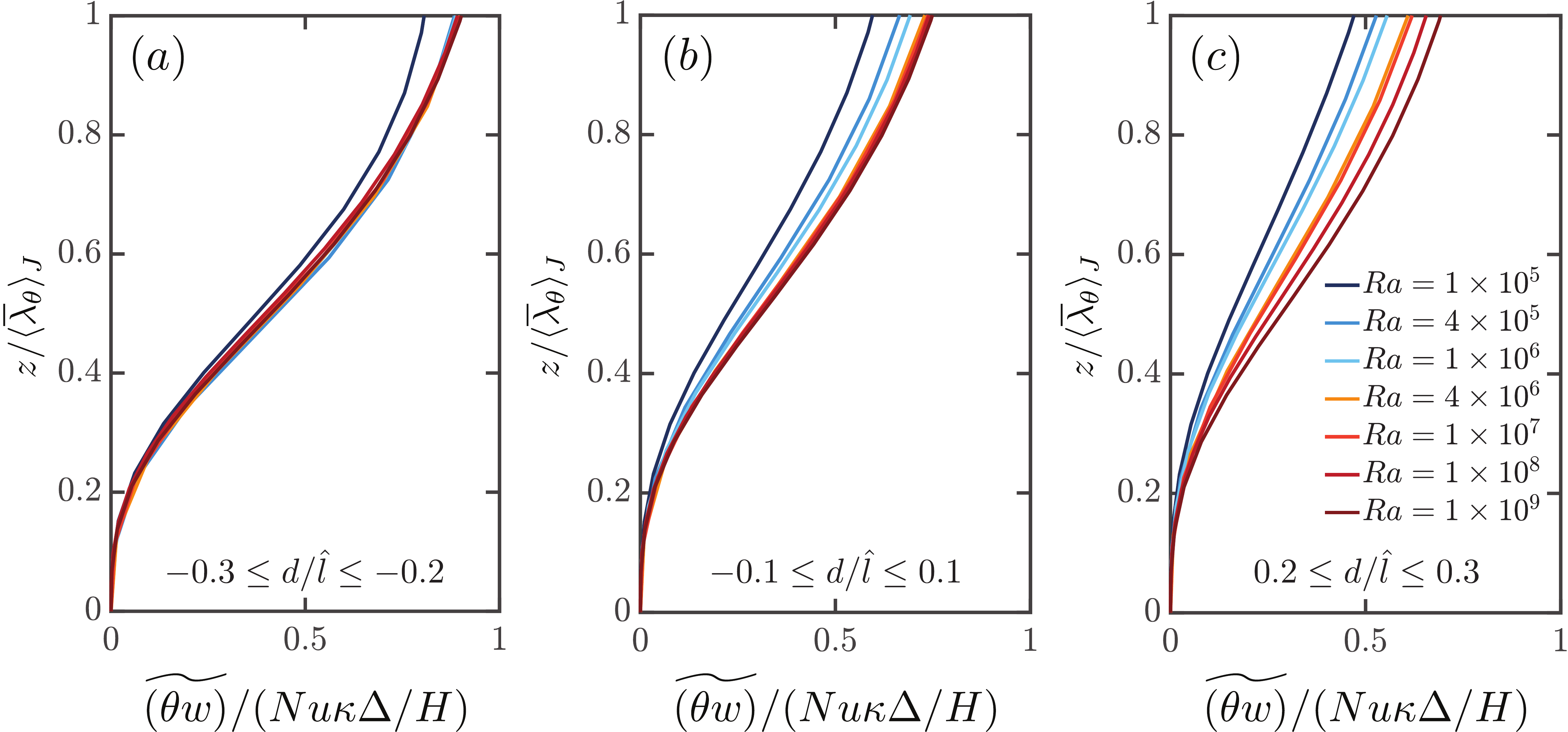}%
 \caption{\label{fig:heattransport} Large-scale turbulent heat transport term 
 $\widetilde{(\theta w)} / ( Nu \kappa \Delta/H)$
  evaluated in (a) the plume impacting region, (b) for small $|d|$ around zero, and (c) in the plume emitting region.}
\end{figure}

Next, we consider the local heat flux at the plate surface, given by 
\begin{equation}
\overline{Nu}(d) \equiv -  {H \over\Delta  }\partial _z \overline{\theta}(d),
\end{equation}
which is plotted in figure \ref{fig:Nu}a for the full range of $Ra$. In all cases $\overline{Nu}/Nu$ is higher than one on the plume impacting side ($d<0$). This is consistent with the impacting cold plume increasing the temperature gradient in the BL locally. The fluid subsequently heats up while it is advected along the plate towards increasing $d$ by the LSC. As a consequence, the wall gradient is reduced and $\overline{Nu}$ decreases approximately linearly with increasing $d/\hat{l}$, which is consistent with observations by \citet{ree08} and \citet{wag12}. This leads to the ratio $\overline{Nu}/Nu$ dropping below 1 for $d>0$. For increasing $Ra$, the local heat flux becomes progressively more uniform across the full range of $d$. To quantify this, we plot the mean local heat fluxes in the plume impacting and emitting regions, respectively, in figure \ref{fig:Nu}b. The former is decreasing while the latter is increasing with increasing $Ra$, bringing the two sides closer. Again, and in both cases, a change of slope is visible in the range of $Ra=1\times 10^6 \textup{--} 4\times 10^6$. In this context it is interesting to note that in a recent study on two-dimensional RB convection \citep{zhu18c} it was found that at significantly higher $Ra \gtrapprox 10^{11}$ the heat transport in the plume emitting range dominated, reversing the current situation. If we extrapolate the trend for $Ra\geq 4\times 10^6$ in our data, we can estimate that a similar reversal may occur at $Ra \approx \mathcal{O}(10^{12}\textup{--}10^{13})$, see figure \ref{fig:Nu}b.

A possible mechanism that might explain this behavior is increased turbulent (or convective) mixing, which can counteract the diffusive growth of the temperature BLs. To check this hypothesis, we plot the heat transport term $\widetilde{(\theta w)} \equiv \overline{w \theta} - \overline{w} \overline{\theta}$ in figure \ref{fig:heattransport}. The normalization in the figure is with respect to the total heat flux $Nu$, 
 the plotted quantity reflects the fraction of $Nu$ carried by $\widetilde{(\theta w)}$. It is obvious from these results that the convective transport contributes significantly, even within the BL height $\langle\overline{\lambda}_\theta \rangle_J$. Moreover, this relative contribution is independent of $Ra$ (except for the lowest value considered) in the plume impacting region (see figure \ref{fig:heattransport}a). However, figure \ref{fig:heattransport}b shows that already around $d = 0 $ the convective transport in the BL increases with increasing $Ra$. This trend is much more pronounced in the plume emitting region $d>0.2$, see figure \ref{fig:heattransport}c. Hence, convective transport in the BL plays an increasingly larger role for $d\geq 0$ with increasing $Ra$. Its effect is to mix out the near-wall region, thereby increasing the temperature gradient at the wall. It is conceivable that the increased convective transport in the near-wall region (provided the trend persists) eventually leads to a reversal of the $\overline{Nu} (d)$ trend observed at moderate $Ra$ in figure \ref{fig:Nu}a.

\section{Thermal and viscous boundary layers}
\label{section_BL}

\begin{figure}
 \centering
 \includegraphics[width=\textwidth]{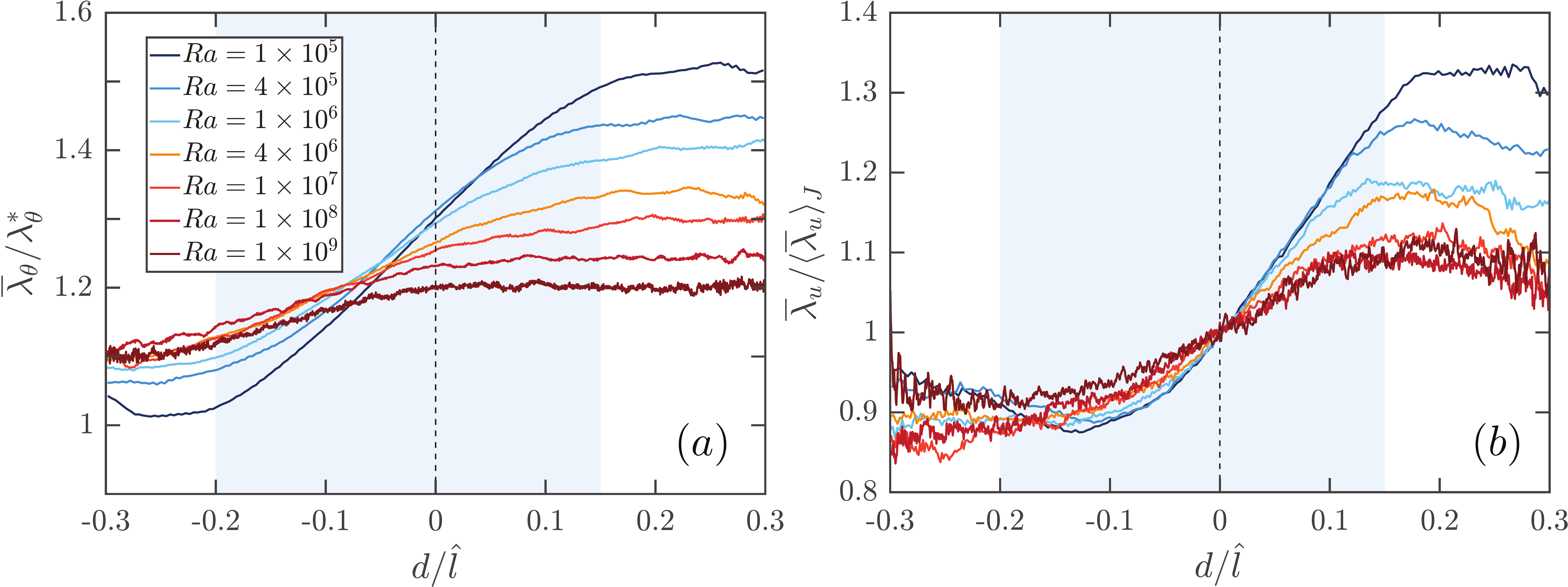}%
 \caption{\label{fig:lambda} (a) Thermal BL thickness $\overline{\lambda} _\theta$ normalized by the estimated thermal BL thickness $\lambda _\theta ^*$ and (b) viscous BL thickness $\overline{\lambda} _u$ normalized by the mean viscous BL thickness in the interval $d/ \hat{l} \in J$ versus normalized distance $d/ \hat{l}$. The color indicates the Rayleigh number, see legend.}
\end{figure}

Next, we study how the BL thicknesses $\lambda_\theta$ and $\lambda_u$ vary along the LSC. In figure \ref{fig:lambda}a we present $\overline{\lambda}_\theta$, normalized by $\lambda_\theta ^*$. As expected from figure \ref{fig:Nu}, $\overline{\lambda}_\theta$ is generally smaller in the plume impacting region and then increases along the LSC. However, unlike $\overline{Nu}$, $\overline{\lambda}_\theta$ is not determined by the gradient alone but also depends on the temperature level (see figure \ref{fig:counts}b) such that differences arise. Specifically, $\overline{\lambda}_\theta/\lambda_\theta ^*$ is rather insensitive for $Ra \geq 4\times 10^6$ in the plume impacting region ($d/\hat{l}<-0.1$). Furthermore, for $Ra \geq 10^7$, the growth of the thermal BL with $d/\hat{l}$ comes to an almost complete stop beyond $d= 0$, which is entirely consistent with the conclusions drawn in the discussion on $\widetilde{(\theta w)}$ above. Finally, we note that $\overline{\lambda}_\theta$ is generally larger than the estimate $\lambda_\theta ^*$, which agrees with previous observations by \cite{wag12}.

\begin{figure}
 \centering
 \includegraphics[width=\textwidth]{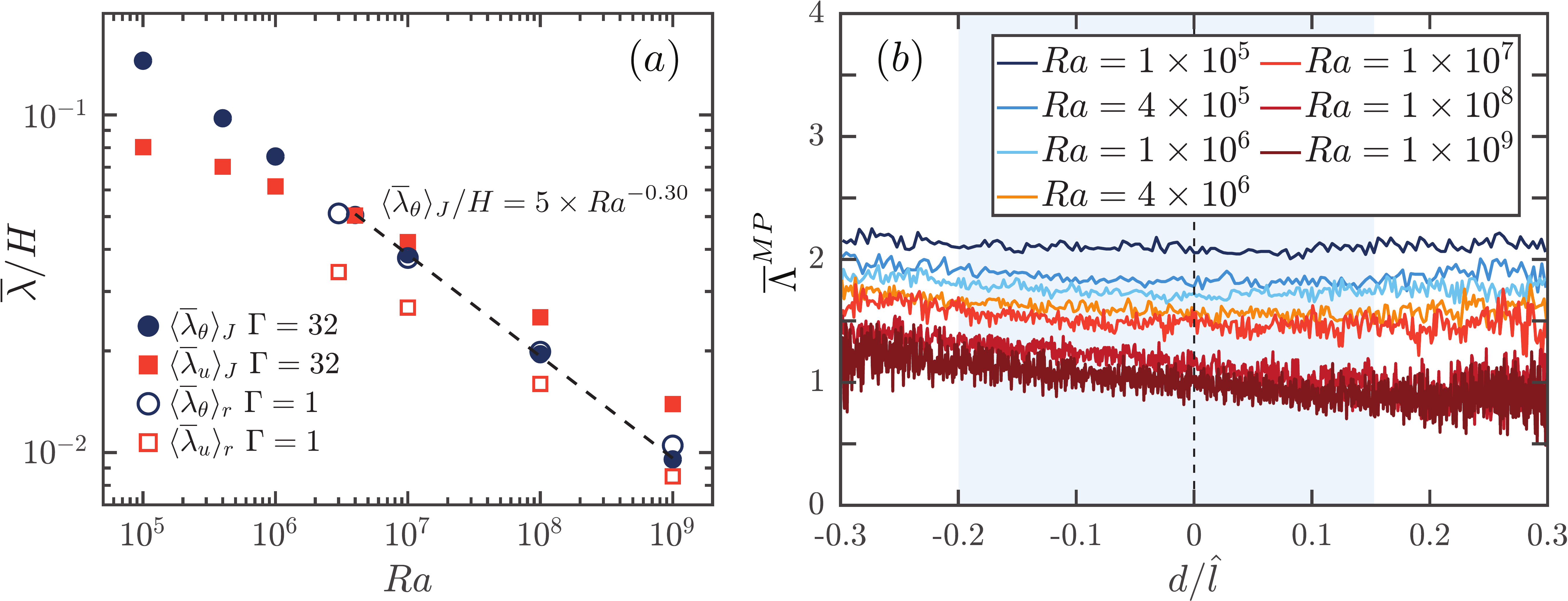}%
 \caption{\label{fig:lambda_tu}(a) Mean BL thicknesses versus $Ra$ for the present data at $\Gamma = 32$ (filled symbols) and those of \citet{wag12} with $Gamma = 1$ (open symbols).  (b) Most probable BL ratio $\overline{\Lambda} ^{MP} = \lambda_\theta / \lambda_u $ versus normalized distance $d/ \hat{l}$ for various $Ra$.}
\end{figure}

There is no obvious choice for the normalization of the viscous BL thickness and we therefore present $\overline{\lambda} _u$ normalized with its mean value $\langle \overline{\lambda} _u \rangle_{J}$ in figure \ref{fig:lambda}. Overall these curves for $\overline{\lambda} _u$ exhibit a similar trend as we observed previously for $\overline{\lambda} _\theta$. The values of $\overline{\lambda} _u$ are smaller in the plume impacting region ($d<0$) and the variation with $Ra$ is limited. Also for $\overline{\lambda} _u/ \langle \overline{\lambda} _u \rangle_{J}$ the growth with increasing $d$ is less pronounced the higher $Ra$ and the curves almost collapse for $d>0$ at $Ra\geq 10^7$. 

Figure \ref{fig:lambda_tu}a shows $\langle \overline{\lambda}_\theta \rangle_{J}$ and $\langle \overline{\lambda}_u \rangle_{J}$ as a function of $Ra$. For the thermal BL thickness, the scaling appears to be rather constant over the full range and from fitting $4 \times 10^6 \leq Ra \leq 10^9$ we obtain
\begin{equation}
\langle \overline{\lambda}_\theta \rangle_{J}/H \sim Ra^{-0.30}.
\end{equation}
The reduction of the viscous BL thickness $\langle \overline{\lambda}_u \rangle_{J}$ with $Ra$ is significantly slower than for the thermal BL thickness $\langle \overline{\lambda}_\theta \rangle_{J}$. For low $Ra$, $\langle \overline{\lambda}_u \rangle_{J} < \langle \overline{\lambda}_\theta \rangle_{J}$. However, due to the different scaling of the two BL thicknesses, $\langle \overline{\lambda}_u \rangle_{J} > \langle \overline{\lambda}_\theta \rangle_{J}$ for $Ra \approx 4\times 10^6$. Comparing the periodic $\Gamma =32$ data with the confined $\Gamma =1$ case reported in \cite{wag12}, we note that the results for $\langle \overline{\lambda}_\theta \rangle$ agree closely between the two geometries. The scaling trends for $\langle \overline{\lambda}_u \rangle$ also appear to be alike in both cases. However, the viscous BL is significantly thinner in the smaller box. This situation is similar, and obviously also related to, the findings we reported for the comparison of the wall-shear stress in figure \ref{fig:tau_w}b.

We further computed the instantaneous BL ratio $\Lambda = \lambda_\theta / \lambda _u$ for which results are presented in figure \ref{fig:lambda_tu}b. Since the statistics of $\Lambda$ were found to be quite susceptible to outliers, we decided to report the most probable value $\overline{\Lambda}^{MP}$ as this provides a more robust measure than the mean. The Prandtl-Blasius BL theory for the flow over a flat plate suggests that $\Lambda = 1$ for $Pr= 1$. The figure shows that $\overline{\Lambda}^{MP}$ is almost constant as function of $d/\hat{l}$. However, unexpectedly, $\overline{\Lambda}^{MP}$ turns out to depend on $Ra$. For $Ra =10^5$, $\overline{\Lambda}^{MP}\approx 2$, which is larger than the theoretical prediction, but similar to the ratio of the means reported in figure \ref{fig:lambda_tu}a. $\overline{\Lambda}^{MP}$ decreases with $Ra$ and approaches the predicted value of $1$ for $Ra=10^9$. We note that, although this $Ra$ dependence is not expected, it was also observed by e.g.\ \cite{wag12}.

\begin{figure}
 \centering
 \includegraphics[width=0.7\textwidth]{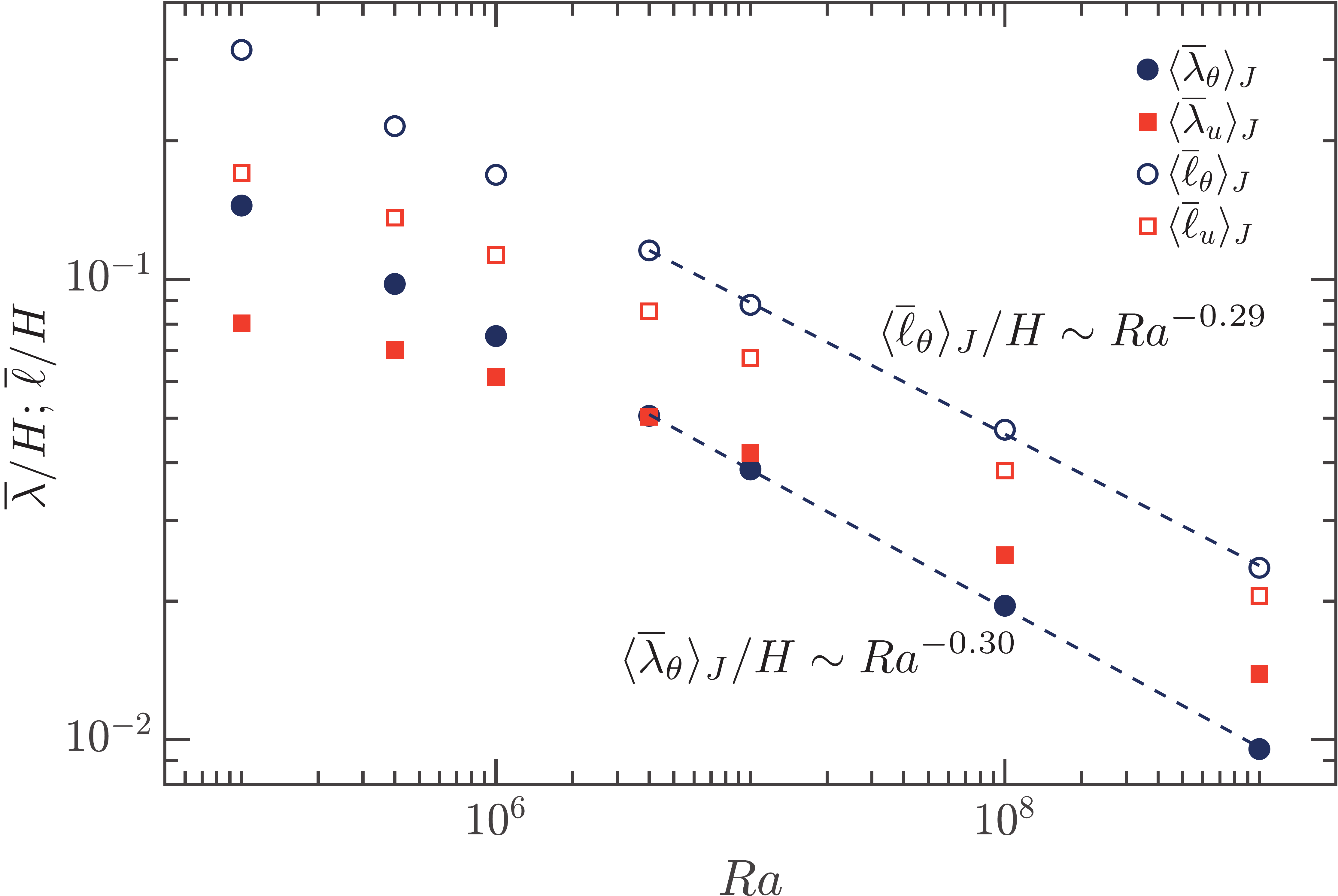}%
 \caption{\label{fig:lambda_ell} Comparison of mean BL thicknesses versus $Ra$ using the slope method and the location of the respective temperature and velocity levels (level method).}
\end{figure}

When interpreting results for the BL thicknesses, it should be kept in mind that different definitions exist in the literature \citep{pui07,zho10,zho10b,sch12,zho13,pui13,sch14,shi15,shi17b,chi19}. We note that values may depend on the boundary layer definition that is employed. To get at least a sense for which of the observations transfer to other possible BL definitions, we compare the results for $\lambda$ (the slope method) to those obtained by the location of the temperature and velocity levels $(\overline{\ell})$ (level method, see figure \ref{fig:counts}b) in figure \ref{fig:lambda_ell}. We note that the scalings versus $Ra$ are very similar, albeit not exactly the same, for both definitions of the BL thickness. However, the offset between $\overline{\lambda}$ and $\overline{\ell}$ is not the same for velocity and temperature. As a consequence, there is no crossover between $\overline{\ell_\theta}$ and $\overline{\ell_u}$ within the range of $Ra$ considered.

\begin{figure}
 \centering
 \includegraphics[width=\textwidth]{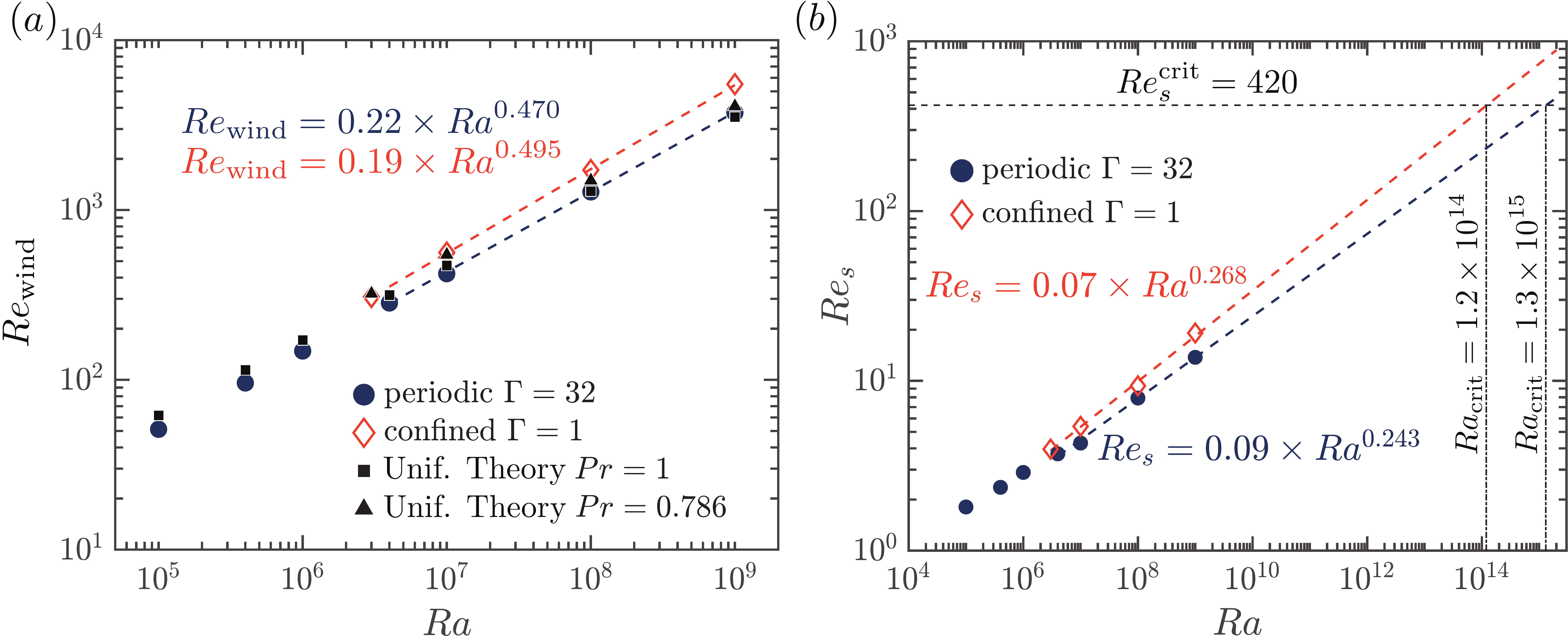}
 \caption{\label{fig:re_res} (a) Wind Reynolds number $Re_\textrm{wind}$ versus $Ra$ obtained in a periodic $\Gamma=32$ domain compared to the corresponding values obtained by \cite{wag12} for a cylindrical $\Gamma =1$ domain. We also show the predictions from the unifying theory \citep{gro00,gro01} using the updated prefactors \citep{ste13}. (b) $Re_s$ versus $Ra$ with estimations for $Ra_\textrm{crit}$. In both panels, we have fitted our own datapoints only from $Ra=4\times 10^6$ onwards to achieve consistent comparisons with the data by \cite{wag12}, where only data from $Ra=3\times 10^6$ on is available.}
\end{figure}

In figure \ref{fig:re_res} we compare the wind Reynolds number, which we determined as follows, 
\begin{equation}
Re_{\textrm{wind}} = \langle \overline{u}_L \rangle_{J} \, H / \nu ,
\end{equation}
with the results of \cite{wag12}. The figure shows that our $Re_{\textrm{wind}}$ obtained in a periodic $\Gamma=32$ domain with $Pr=1$ agree surprisingly well with the results from \cite{wag12} obtained in a cylindrical $\Gamma=1$ sample with $Pr=0.786$. The obtained $Re$ values obtained by \cite{wag12} are slightly higher than our values. We note that the lower $Pr$ results in slightly higher $Re_\textrm{wind}$. This means that the main finding in this context is that $Re_\textrm{wind}$ in the turbulent superstructures is almost the same, perhaps slightly lower, than in a confined $\Gamma=1$ sample \citep{wag12}. We note that the predictions for the wind Reynolds number obtained from the unifying theory for thermal convection \citep{gro00,gro01} are in good agreement with the data. The unifying theory, using the updated constants found by \cite{ste13}, namely predicts that for $Pr=1$ the wind Reynolds number scales as
%
$Re_{GL}=0.395\times Ra^{0.439}$,
%
while the data for $Ra \geq 4\times10^6$ are well approximated by
%
$Re_{\textrm{wind}}=0.22 \times Ra^{0.470}$.

To estimate when the BLs become turbulent we calculate the shear Reynolds number 
\begin{equation}
Re_s=   { \left [ \overline{u}_L \times \lambda _u^{MP} \right ] ^{\max}    \over 2 \nu}.
\end{equation}
We expect the BL to become turbulent and the ultimate regime to set in \citep{gro00,gro11} at a critical shear Reynolds number of $Re_s^\textrm{crit} \approx 420$ \citep{ll87}. A fit to our data gives
\begin{equation}
Re_s=0.09\times Ra^{0.243},
\end{equation}
from which we can extrapolate that $Re_s^\textrm{crit}=420$ is reached at $Ra_\textrm{crit} \approx 1.3\times 10^{15}$. Of course, this estimate comes with a significant error bar as our data for $\Gamma =32$ is still far away from the expected critical $Ra$ number. Nevertheless, it agrees well with the result from \cite{wag12}, who find $Ra_\textrm{crit}\approx 1.2\times 10^{14}$ for a cylindrical $\Gamma=1$ cell and the results from \cite{sun08} who find from experiments that $Ra_\textrm{crit}\approx 2\times 10^{13}$. We emphasize that all these estimates are consistent with the observation of the onset of the ultimate regime at $Ra_{*}\approx 2\times 10^{13}$ in the G\"ottingen experiments \citep{he12,he15}. As is explained by \cite{ahl17} also measurements of the shear Reynolds number in low $Pr$ number simulations by \cite{sch16} support the observation of the ultimate regime in the  G\"ottingen experiments.

\section{Conclusions} \label{section_conclusions}
We have used a conditional averaging technique to investigate the properties of the LSC and the boundary layers in $\Gamma =32$ RB convection for unit Prandtl number and Rayleigh numbers up to $Ra=10^9$. The resulting quasi-two-dimensional representation of the LSC allowed us to analyze the wind properties as well as wall shear and local heat transfer. We found the distribution of the wall shear stress  $\overline{\tau} _w$ to be asymmetric. The maximum of $\overline{\tau} _w$ is located closer to the plume impacting side and its value increases as $\overline{\tau} _w ^{\max} 
/ (\rho \beta g H \Delta )
\sim Ra^{0.24}$ with increasing $Ra$. The local heat transfer at the wall, represented by the conditioned Nusselt number $\overline{Nu}$, has its highest values in the plume impacting zone at all $Ra$ considered here. Going from the plume impacting towards the plume emitting region, $\overline{Nu}$ is seen to decrease consistently as is expected from the fluid near the hot wall heating up. However, as $Ra$ is increased, the differences in $\overline{Nu}$ even out more and more. For the plume emitting side in particular, we were able to connect this trend to increased advective transport in the wall-normal direction at higher $Ra$. When extrapolating the trends for $\overline{Nu}$ to $Ra$ higher than those available here, our results appear consistent with \citet{zhu18c}. These authors observed a reversal of the $\overline{Nu}$-distribution in 2D RB turbulence above $Ra \gtrapprox 10^{11}$ with higher values of the heat transport in the emitting region.

Further, we examined the thermal and the viscous BLs. At low $Ra$, both increase along $d$ in an approximately linear fashion, whereas flat plate boundary layer theory would suggest a growth proportional $\sqrt{d}$ \citep{ll87}. As $Ra$ increases, and especially for $d>0$, the growth becomes successively weaker and stops entirely beyond $Ra \gtrapprox10^8$. Again, this is likely a consequence of the increased convective mixing in this region. For increasing $Ra$, both $\overline{\lambda}_{\theta}$ and $\overline{\lambda}_u$ become thinner, with $\overline{\lambda} _\theta$ showing an effective scaling of $\langle \overline{\lambda} _{\theta} \rangle _J /H \sim Ra^{-0.3}$. At $Ra\gtrapprox 4\times 10^6$ we observed a crossover point where the thermal BL becomes smaller than the viscous BL. It should be noted that the crossover appears specific to the definition of $\overline{\lambda}$ since a similar behavior was not observed when an alternative definition ($\overline{\ell}$, based on the location of the level) was employed. Nevertheless, the scaling behavior of  $\overline{\lambda}$ and $\overline{\ell}$ was seen to be very similar.
When calculating instantaneous BL ratios, a convergence to $\overline{\Lambda}^{MP} \rightarrow 1$ for high enough $Ra$ can be observed as predicted by the PB theory for laminar BLs. As pointed out in \citet{shi14}, the PB limit only strictly applies to wall parallel flow and the ratio is expected to be higher if the flow approaches the plate at an angle. This incidence angle is higher at smaller $\Gamma$ which can explain why at comparable $Ra$ the BL ratios reported in \citet{wag12} are slightly higher than what is found here.

We expected to find significant differences in the LSC statistics obtained in a confined $\Gamma =1$ system and a large $\Gamma =32$ system. However, surprisingly, we find that the thermal BL thickness $\langle \overline{\lambda}_{\theta} \rangle _J$ obtained for both cases agrees very well. It turns out that the viscous BL thickness $\langle \overline{\lambda}_u \rangle _J$ is significantly larger for the periodic $\Gamma =32$ case than in a $\Gamma=1$ cylinder. However, the wall shear and its scaling with $Ra$ are similar in both cases. Here we find that in a periodic $\Gamma=32$ domain, the shear Reynolds number scales as $Re_s \sim Ra^{0.243}$. This is a bit lower than the corresponding result for $\Gamma=1$, although one needs to keep in mind the slight difference in $Pr$ ($Pr = 0.786$ at $\Gamma = 1$ vs. $Pr = 1$ for $\Gamma=32$) is responsible for part of the observed difference. An extrapolation towards the critical shear Reynolds number of $Re_s^{\textrm{crit}} \approx 420$ when the laminar-type BL becomes turbulent predicts that the transition to the ultimate regime is expected at $Ra_{\textrm{crit}} \approx \mathcal{O}(10^{15})$. This is slightly higher than the corresponding result for a $\Gamma=1$ cylinder, i.e. $Ra_{\textrm{crit}} \approx \mathcal{O}(10^{14})$, by \citep{wag12}. However, it should be noted that considering inherent uncertainties and differences in $Pr$, the results for $\Gamma=32$ the observed transition to the ultimate regime in the G\"ottingen experiments \citep{he12,he15} and previous measurements of the shear Reynolds number \cite{wag12,sch16}. So surprisingly, we find that in essentially unconfined very large aspect ratio systems, in which the resulting structure size is significantly larger, the differences in terms of $Re_{\textrm{wind}}$ or $Re_s$ with respect to the $\Gamma=1$ cylindrical case are marginal.

\section*{Acknowledgments}
We greatly appreciate valuable discussions with Olga Shishkina. This work is supported by NWO, the University of Twente Max-Planck Center for Complex Fluid Dynamics, the German Science Foundation (DFG) via program SSP 1881, and the ERC (the European Research Council) Starting Grant No. 804283 UltimateRB. The authors gratefully acknowledge the Gauss Centre for Supercomputing e.V. ({\color{blue}\url{www.gauss-centre.eu}}) for funding this project by providing computing time on the GCS Supercomputer SuperMUC-NG at Leibniz Supercomputing Centre ({\color{blue}\url{www.lrz.de}}).

\section*{Declaration of Interests}
The authors report no conflict of interest.

\bibliographystyle{jfm}
\bibliography{literature_turbulence}
\end{document}